\documentclass[twocolumn]{aastex7}

\usepackage{graphicx}
\usepackage{xcolor}
\usepackage{soul}

\def\kms{km$\ $s$^{-1}$}
\def\flam{${\rm erg\; cm^{-2}\; s^{-1}\; \AA^{-1}}$}

\accepted{for publication in The Astrophysical Journal on July 22, 2025.}

\begin{document}

\title{$HST$ Observations of the CV Propeller LAMOST~J024048.51+195226.9}

\author[0009-0003-0467-4440]{Jordan Tweddale}
\affiliation{Department of Physics and Astronomy \\
University of Notre Dame \\ Notre Dame, IN 46556, USA}
\email{jtweddal@nd.edu}

\author[0000-0003-4069-2817]{Peter Garnavich}
\affiliation{Department of Physics and Astronomy \\
University of Notre Dame \\ Notre Dame, IN 46556, USA}
\email{pgarnavi@nd.edu}

\author[0000-0001-7746-5795]{Colin Littlefield}
\affiliation{Bay Area Environmental Research Institute, Moffett Field, CA 94035 USA}
\email{clittlef@alumni.nd.edu}

\author[0000-0003-4373-7777]{Paula Szkody}
\affiliation{Department of Astronomy, University of Washington, Seattle, WA 98195, USA}
\email{szkody@uw.edu}

\begin{abstract}

We present \textit{Hubble Space Telescope} ($HST$) FUV spectra and light curves of the magnetic cataclysmic variable (CV) LAMOST~J024048.51+195226.9 (J0240), the second known CV propeller. The five consecutive $HST$ orbits span a full 7.34 hour binary orbital period. We detect a 24.939~$\pm$~0.006~s FUV modulation, confirming that J0240 contains the fastest spinning white dwarf (WD) in a CV. A high \ion{N}{5}/\ion{C}{4} emission line ratio is considered an indicator of a recent episode of thermal time-scale mass transfer. The observed ratio in J0240 is higher than seen in typical magnetic CVs, but far less than observed in the only other confirmed propeller, AE Aquarii (AE Aqr). We also find that J0240 is significantly less luminous than AE Aqr during both low- and high-flux states. Around orbital phase 0.5, the \ion{Si}{4} emission line displays a P-Cygni absorption profile likely related to the gas accelerated in the propeller. We derive new mass-dependent temperature limits for the surface temperature of the WD of T~$\leq$~11,000-15,000 K. This temperature is low enough to allow for WD core crystallization, which may be linked to magnetism in WDs, particularly those in CVs.

\end{abstract}


\section{Introduction} \label{sec:intro}

Cataclysmic variables (CVs) are binary star systems composed of a white dwarf (WD) primary and a cool, sub-solar mass, secondary ``donor" star. The two stars are very tightly bound; normal orbital periods range from 80~minutes to 5 hours, with a period gap between 2 and 3~hours \citep{Inight2023}. The red dwarf fills its Roche lobe, allowing gas to flow through the gravitationally stable first Lagrange point, L1. Typically, donated gas will form an accretion disk around the WD, where matter loses angular momentum as it gradually spirals onto the surface of the WD \citep[see][for a review]{Warner2003}. 

In some cases, the WD in a CV has an appreciable magnetic field, altering the flow of material toward the WD. These CVs are separated into two classes: polars and intermediate polars (IPs). Polars have magnetic fields in excess of 10~MegaGauss (MG), a field that is generally strong enough to prevent an accretion disk from forming, and causes the polar to rotate synchronously or very slightly asynchronously. The lower magnetic fields ($\sim$~1-10~MG) found in IPs result in more complex accretion processes and WD spin periods that are typically significantly faster than the binary orbital period \citep{King1991}. See \citet{cropper90} and \citet{Patterson1994} for reviews on polars and IPs.

In extreme cases, a CV with a rapidly spinning magnetic WD can accelerate and eject the secondary's overflow from the binary system \citep{Wynn1997}. These systems are known as ``magnetic CV propellers". The WD in a CV propeller is believed to have been spun-up to a high rate of rotation during a recent episode of thermal-timescale mass transfer from the secondary \citep{Schenker2002}. A normal accretion disk is prevented from forming and donated gas is ejected in the plane of the binary orbit. These are distinct from propellers in X-ray binaries, which possess disks that interact with the magnetosphere of a spinning neutron star, and ejected gas is generally directed out of the disk plane \citep{burderi1996}.

For decades, AE~Aquarii (AE~Aqr) was the only known member of the magnetic CV propeller subclass. AE~Aqr has an extremely short WD spin period of 33~s, and an unusually long orbital period of 9.88~hours \citep{Patterson1979}, but is perhaps most notable for exhibiting large-amplitude flares that have been observed across radio, optical, UV, and X-ray wavelengths \citep[see respectively][]{Abada-Simon1993,Patterson1979,Eracleous1994,Osborne1995}. These flares last for several minutes and occur multiple times per day, but not predictably. The flares in AE~Aqr have amplitudes of up to one magnitude at blue wavelengths. Furthermore, AE Aqr appears not to possess an accretion disk and its WD spin period has been observed to be increasing at a rate of 5.64x10$^{-14}$~s~s$^{-1}$ \citep[spin-down;][]{deJager1994}. This suggests that the WD rotation is supplying the energy to eject gas from the system, and that the propeller state is a short-lived epoch ($\sim$ 10$^{7}$ years) in the evolution of the system. The peculiar observational properties of AE Aqr have been attributed to the overflow of diamagnetic gas blobs from the secondary that subsequently interact with the WD's rapidly rotating magnetic field \citep[see][]{King1993,Wynn1995,Eracleous+Horne1996,Wynn1997}.
 
AE~Aqr recently gained some company. LAMOST J024048.51+195226.9 (hereafter J0240) was originally classified as a CV by its emission line spectrum \citep{Hou2020}. Follow-up optical photometry revealed frequent flaring events \citep{Thorstensen2020}, which led to J0240 being reclassified as the second confirmed CV propeller \citep{Garnavich2021}. The WD in J0240 possesses a 24.93~s spin period, making it the fastest known spinning WD within a CV system \citep{Pelisoli2022}. Like AE~Aqr, J0240 has a relatively long orbital period (7.34 hours) compared with other CVs \citep{Littlefield2020}. One key difference between J0240 and AE~Aqr is that the orbital plane of J0240 is viewed nearly edge-on and exhibits eclipses of the optical flaring region \citep{Littlefield2020}. From the eclipse timing, \citet{Garnavich2021} concluded that the region generating the continuum flares is very close to the WD. The high inclination also results in narrow P-Cygni absorption features, likely coming from gas that has been ejected by the propeller mechanism \citep{Garnavich2021}.

We present the first FUV study of J0240 using time-resolved \textit{Hubble Space Telescope} ($HST$) Cosmic Origins Spectrograph (COS) UV spectroscopy. We compare its properties with those of AE~Aqr studied by \citet{Eracleous+Horne1996}. Specifically, we search for the UV spin modulation of J0240's WD and estimate the \ion{C}{4} emission line ratios, which have been observed to be anomalous in CVs that host extremely rapidly rotating WDs.

\section{Observations} \label{sec:obs}

\subsection{\textit{HST}}

J0240 was observed by $HST$-COS over 5 contiguous $HST$ orbits on 2022-02-03 (GO-16881, PI: Garnavich). These visits produced 6 exposures totaling 13062~s, with the fourth $HST$ orbit being divided into two consecutive exposures. Adopting the ephemeris presented in \citet{Garnavich2021}, the 5 $HST$ orbits span $-0.050\leq\phi\leq0.904$ of the 7.33 hour binary orbital period, where $\phi$ is the binary orbital phase, and $\phi=0$ corresponds to the eclipse of the WD by the secondary. These observations were taken with the G140L grating in TIME-TAG mode, which records individual photons and allows for detailed time-resolved spectroscopic analysis. The FUV channel (segment A) was used in the 1105~{\AA} central wavelength setting. The observations cycled through all four FP-POS positions, giving a total wavelength coverage of 1101-2291~{\AA} at a resolution of R{$\sim$}2000. Due to the low photon count rate long-ward of 2000~{\AA} and the shifting FP-POS positions, the effective wavelength coverage of the spectroscopy was 1150-2000~{\AA}. Information detailing each exposure is presented in \autoref{tab:HSTobs_info}. The data from the COS observations are publicly available through the Mikulski Archive for Space Telescopes (MAST), DOI:\dataset[10.17909/7aw3-ev46]{http://dx.doi.org/10.17909/7aw3-ev46}.

\begin{deluxetable*}{ccccccc}
    \tablecaption{$HST$-COS observations of J0240 }

    \tablenum{1}

    \tablehead{\colhead{Exp ID} & \colhead{$HST$ Orbit} & \colhead{Exp Start} & \colhead{Exp Time} & \colhead{FP-POS} & \colhead{$\lambda$ Range} & \colhead{Orbital Phase ($\phi$)} \\ 
    \colhead{} & \colhead{} & \colhead{(UTC, Feb 3, 2022)} & \colhead{(s)} & \colhead{} & \colhead{({\AA})} & \colhead{} } 

    \startdata
        lesg01r2q &   1 &   01:41:45 &   2376 &   1 &   1158-2291 &   -0.05-0.04 \\ 
        lesg01r4q &   2 &   03:10:48 &   2700 &   2 &   1139-2273 &    0.15-0.26 \\ 
        lesg01r6q &   3 &   04:46:06 &   2700 &   3 &   1118-2251 &    0.37-0.47 \\ 
        lesg01rcq &   4 &   06:21:23 &   1285 &   3 &   1118-2251 &    0.59-0.63 \\ 
        lesg01s1q &   4 &   06:44:43 &   1300 &   4 &   1101-2234 &    0.63-0.69 \\ 
        lesg01snq &   5 &   07:56:40 &   2700 &   4 &   1101-2234 &    0.80-0.90 \\ 
    \enddata

\end{deluxetable*}
\label{tab:HSTobs_info}

\subsection{MDM}

Time-resolved photometry of J0240 was obtained with the 1.3-m McGraw-Hill Telescope at the MDM Observatory. Observations began on 2022 February 3.12 (UT) and ended after 2.8 hours. The photometry overlapped with the second and third $HST$ orbits, providing simultaneous comparison between the FUV and optical variability. Exposures were 30~s in length through a Johnson B-band filter. The Templeton CCD readout section was set to 256$\times$256 pixels to reduce overheads. A total of 309 observations with a cadence of 32.8~s was obtained.

The images were bias subtracted and flat-field corrected. Aperture photometery was performed on J0240 and a comparison star 60~arcseconds to the northeast. The comparison star has a non-variable B-band magnitude of 16.00~$\pm$~0.07~mag \footnote{From the APASS catalog \citep{Henden2019}}. 

\begin{figure*}[ht!]
\plotone{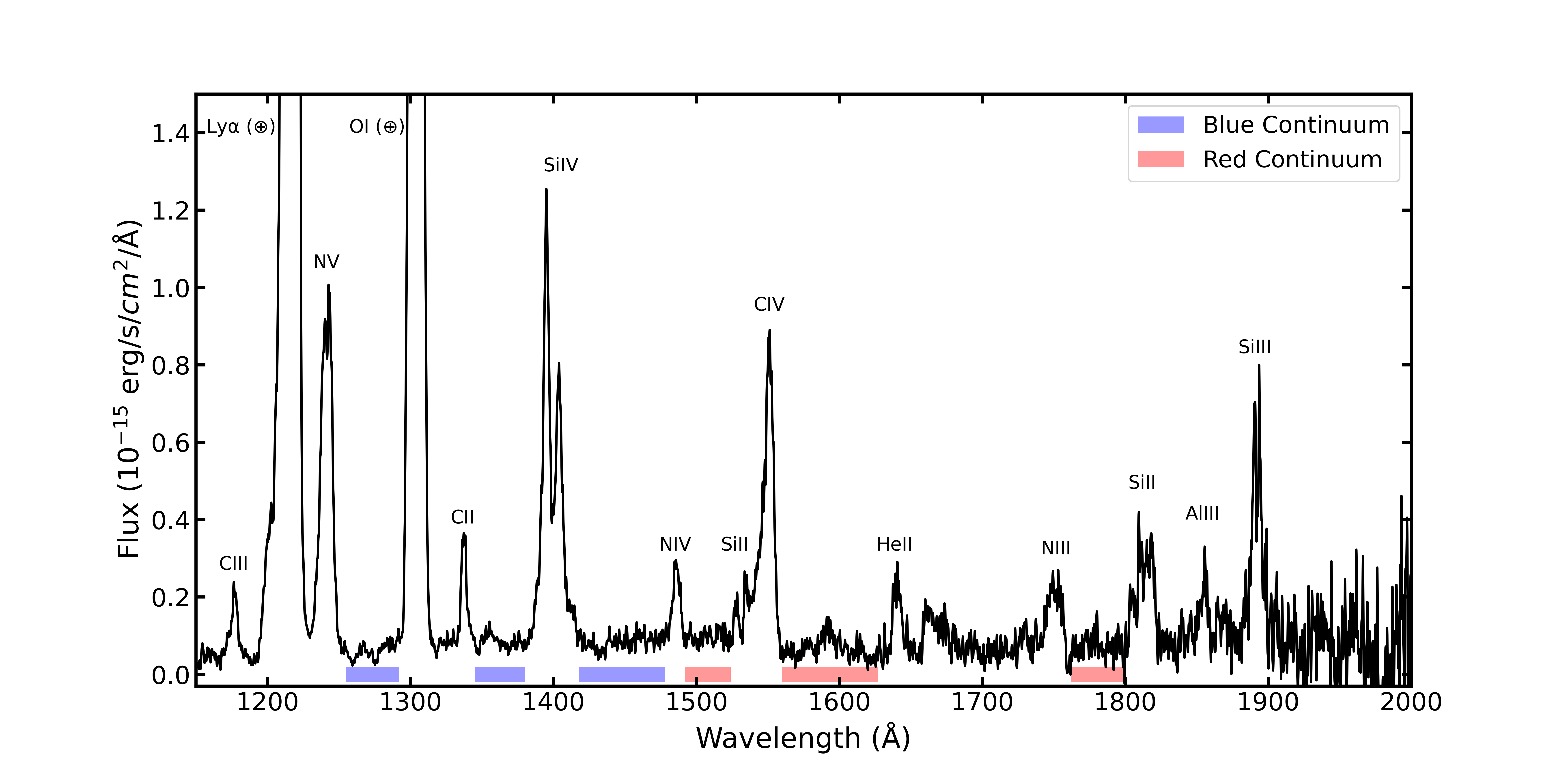}
\caption{The mean UV spectrum of all 5 $HST$ orbits over our effective wavelength coverage of 1150-2000~{\AA}. The blue continuum bands (1255-1292, 1345-1380, 1418-1478~{\AA}) and red continuum bands (1492-1524, 1560-1627, 1762-1800~{\AA}) are highlighted. Line identifications are taken from \citet{Eracleous+Horne1996}.
\label{fig:hst_spec}}
\end{figure*}

\section{Analysis} \label{sec:analysis}

\subsection{Mean {\it HST} Spectrum}

The average FUV J0240 spectrum is a combination of all 5 $HST$ orbits from the \textit{x1dsum} file, which is calibrated through the COS pipeline (calcos). Spectra from individual orbits were analyzed using the calcos calibrated and exposure specific \textit{x1d} files. The mean spectrum of all 5 orbits is shown in \autoref{fig:hst_spec}.

The mean spectrum reveals a flat continuum with emission lines typically found in the FUV spectra of CVs. Resonance doublet line of \ion{N}{5}, \ion{Si}{4}, and \ion{C}{4} are strong. Weaker lines from lower ionization states of nitrogen, silicon, and carbon are also present. \ion{He}{2} at 1640~{\AA} is rather weak, but clearly detected. Except for the presence of carbon emission in J0240, the overall FUV spectrum is similar to that of AE~Aqr \citep{Eracleous1994}.

\subsection{FUV Light Curves}

The observation's \textit{corrtag} files, which contain the arrival times of individual photons, were broken into 30~s sub-exposures with the use of the COS \textit{splittag} module. These sub-exposures were then run through the \textit{CalCOS} software, which processed the COS data to create a calibrated \textit{x1d} spectrum for each sub-exposure. Wavelength masks from 1205-1226~{\AA} and 1293-1314~{\AA} were applied to the sub-exposure spectra to block out telluric Ly-$\alpha$ and \ion{O}{1} emission. 

\begin{figure*}[ht!]
\plotone{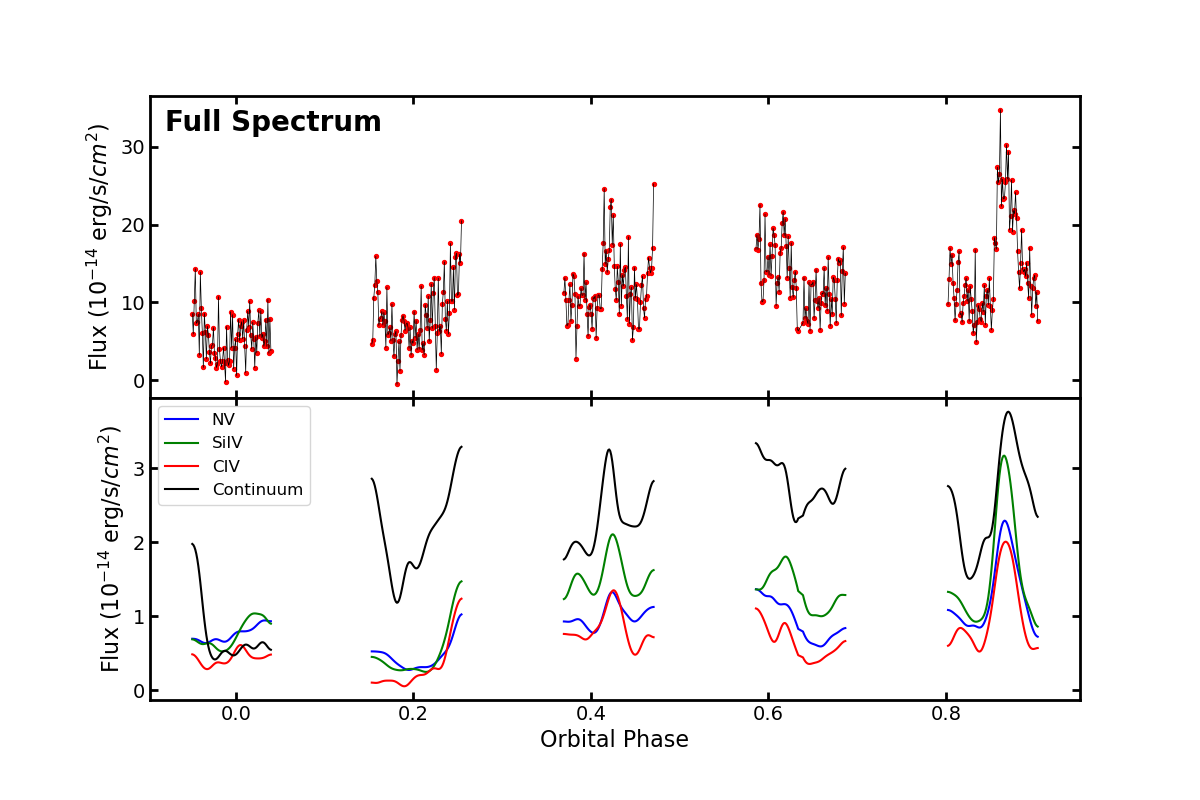}
\caption{The $HST$ light curve of J0240. Fluxes are measured in 30~s bins. The ``Full Spectrum" is defined by the wavelength range 1150-2000~{\AA}, with telluric Ly-$\alpha$ and \ion{O}{1} emission filtered out. The continuum is defined by the sum of the blue and red continuum bands. In the bottom panel, the 30~s bins have been smoothed with a Gaussian filter.
\label{fig:LightCurve_Flux_30s}}
\end{figure*}

The light curve created from the total FUV spectrum with telluric emission removed is shown in the top panel of \autoref{fig:LightCurve_Flux_30s}. The average brightness in the FUV is seen to increase over the $HST$ visit with significant flares detected at the end of orbit 2 ($\phi=0.22$), in orbit~3 ($\phi=0.42$), and in orbit~5 ($\phi=0.86$).

\subsection{Emission Line and Continuum Variations} \label{Emission Line Strengths}

To estimate emission line and continuum variability while maintaining a significant signal-to-noise ratio (SNR), each of the 5 $HST$ orbits were broken into sub-exposures of approximately 400~s in length, resulting in 30 total sub-exposures. This was again accomplished using the \textit{splittag} module with \textit{CalCOS} calibrations. Emission line strengths were calculated for the following ions: \ion{C}{3}~(1175-1176~{\AA}), \ion{N}{5}~(1239,~1243~{\AA}), \ion{C}{2}~(1335~{\AA}), \ion{Si}{4}~(1393,~1403~{\AA}), \ion{N}{4}~(1486~{\AA}), \ion{C}{4}~(1548,~1550~{\AA}), \ion{He}{2}~(1640~{\AA}), and \ion{Si}{3}~(1892~{\AA}). 

A local continuum for each emission line was calculated by integrating a nearby range of wavelength devoid of any obvious spectral features. In most cases, the local continuum was integrated over an equal range of the blue and red sides of an emission line. However, in the case of \ion{N}{5}, only the red side was used to calculate the local continuum due to the presence of the nearby geocoronal Ly-$\alpha$ line. 

The fluxes for each ion were integrated over an appropriate range of ion-specific wavelengths and the local continuum was subtracted. There is a \ion{Si}{2} doublet (1527,~1533~{\AA}) near the blue edge of the \ion{C}{4} line. The 1527~{\AA} \ion{Si}{2} line has a half-width of about 4~{\AA}, so the blue edge of the \ion{C}{4} line is conservatively defined to be 1541~{\AA} to reduce possible contamination from the 1533~{\AA} \ion{Si}{2}. Statistical uncertainties in emission line strengths were estimated from the number of detected photons in each sub-exposure. Systematic uncertainties were estimated by measuring the variations in the line flux resulting from small shifts in the wavelength window used in the integration.

A `total' continuum strength was estimated by summing the flux over six regions in the FUV spectrum that lacked bright emission features. These bands are shown in \autoref{fig:hst_spec}. 
The ``blue continuum" is calculated by integrating the fluxes in the three shortest wavelength bands: 1255-1292, 1345-1380, and 1418-1478~{\AA}. The ``red continuum" is similarly calculated from three longer wavelength bands: 1492-1524, 1560-1627, and 1762-1800~{\AA}. Uncertainties for the blue and red continuum strengths are calculated solely from the statistical uncertainty of the counts in these bands, the total continuum uncertainty is calculated by summing the blue and red continuum uncertainties in quadrature. 

Light curves of the continuum, \ion{N}{5}, \ion{Si}{4}, and \ion{C}{4} fluxes are shown in the bottom panel of \autoref{fig:LightCurve_Flux_30s}. The lowest continuum flux is seen over $HST$ orbit~1, which corresponds to inferior conjunction, where the secondary is between the WD and the observer. The emission lines are found to be their faintest during orbit~2 around $\phi=0.2 $. The flares are detected in both the continuum and line emission.

\subsection{Line Ratio Variations}

\begin{figure}
\plotone{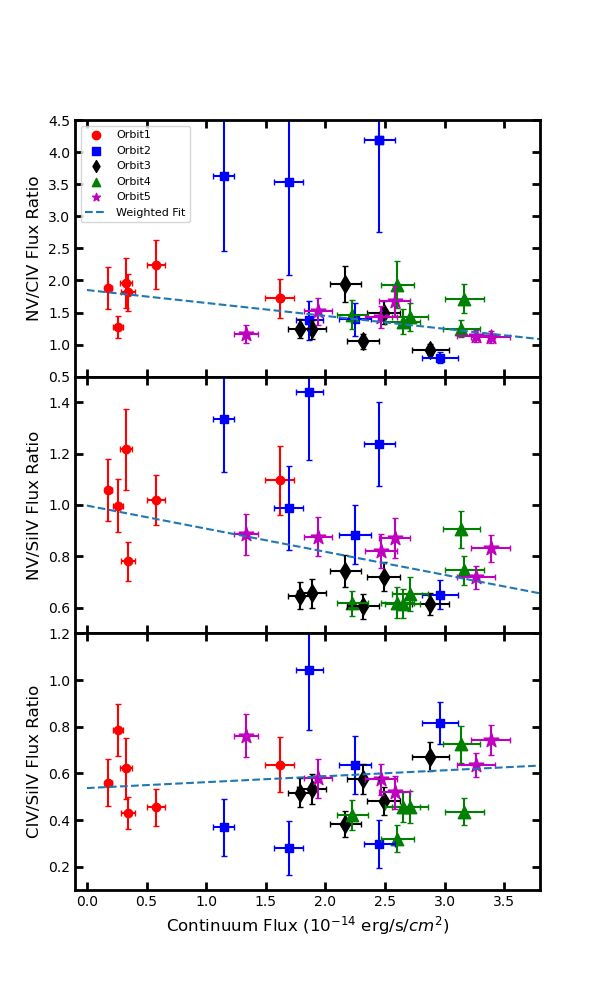}

\caption{  \ion{N}{5}/\ion{C}{4} (top panel), \ion{N}{5}/\ion{Si}{4} (middle panel), and \ion{C}{4}/\ion{Si}{4} (bottom panel) ratios vs. continuum flux. A weighted linear fit shows a weak negative correlation between both \ion{N}{5}/\ion{C}{4} vs. continuum and \ion{N}{5}/\ion{Si}{4} vs. continuum. There is no significant correlation between \ion{C}{4}/\ion{Si}{4} vs. continuum.
\label{fig:EmissionRatios_Continuum}}
\end{figure}

The emission line flux ratios for the three brightest emission lines (\ion{N}{5}, \ion{Si}{4}, and \ion{C}{4}) were calculated and the \ion{N}{5}/\ion{C}{4}, \ion{N}{5}/\ion{Si}{4}, and \ion{C}{4}/\ion{Si}{4} vs. continuum are displayed in \autoref{fig:EmissionRatios_Continuum}. Applying a weighted linear fit to the line ratio vs. continuum flux data suggests a weak relation where the \ion{N}{5}/\ion{C}{4} and \ion{N}{5}/\ion{Si}{4} ratios decrease when the continuum brightens. There is no apparent relation between the \ion{C}{4}/\ion{Si}{4} ratio and continuum flux. To calculate the significance of these correlations we utilized weighted Pearson r-coefficients along with their associated p-values. The (r, p) significance of the linear correlation for \ion{N}{5}/\ion{C}{4} is (-0.37, 0.045); \ion{N}{5}/\ion{Si}{4} is (-0.55, 0.0015); and \ion{C}{4}/\ion{Si}{4} is (0.034, 0.86). Weights are calculated as the inverse of the uncertainty in the emission line ratio.

\begin{figure}
\plotone{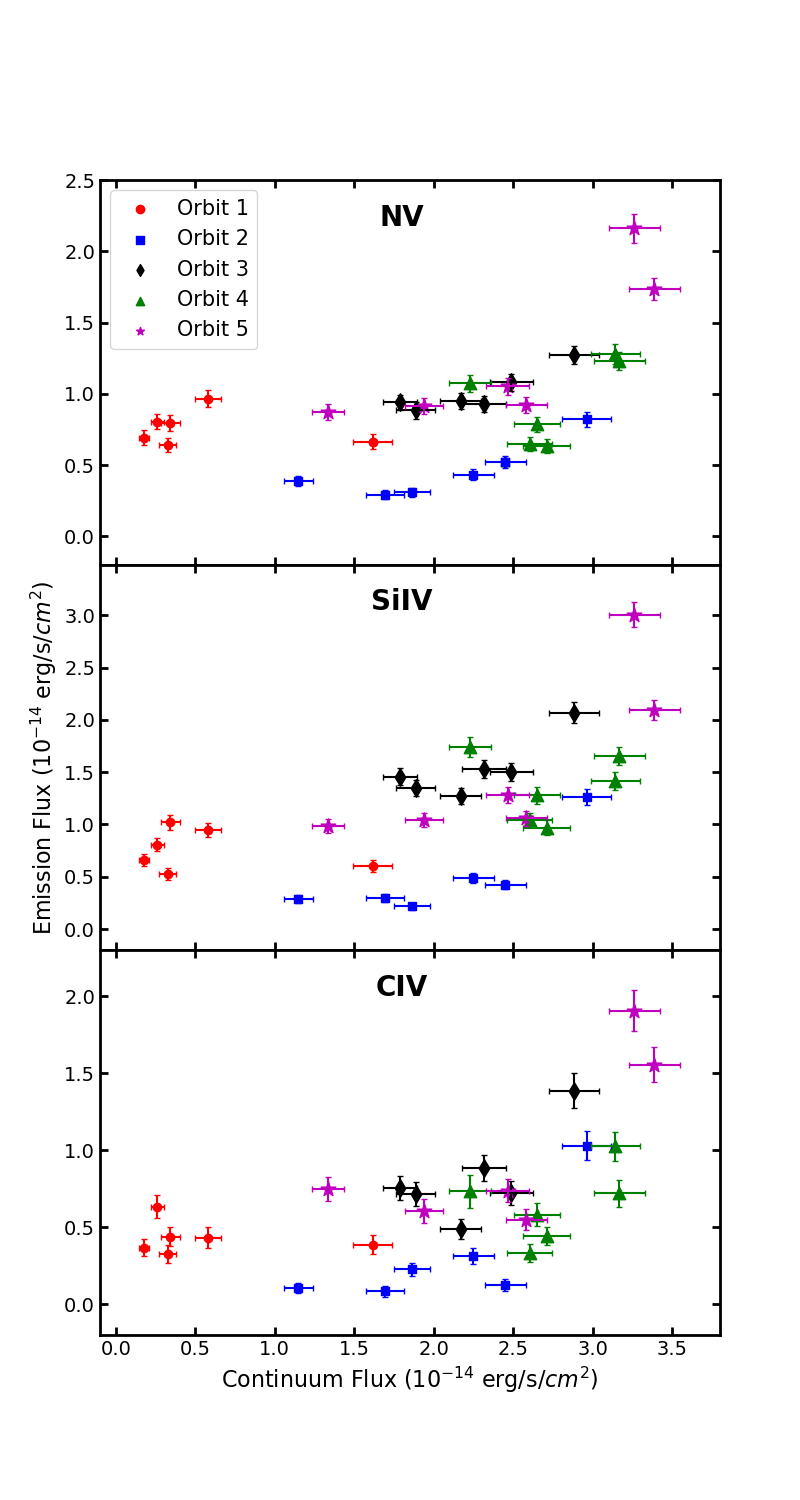}
\caption{ Emission line flux vs. continuum flux for \ion{N}{5} (top panel), \ion{Si}{4} (middle panel), and \ion{C}{4} (bottom panel). Emission line flux remains roughly constant at intermediate continuum strengths, but increases dramatically during flaring events. The emission line flux is strong relative to the continuum during the eclipse (orbit~1), and low relative to the continuum during the first part of orbit~2 ($0.15\leq\phi\leq0.18$; see \autoref{fig:LightCurve_Flux_30s}).
\label{fig:Flux_Continuum}}
\end{figure}

The \ion{N}{5}, \ion{C}{4}, and \ion{Si}{4} line strengths are also plotted versus the continuum flux in \autoref{fig:Flux_Continuum}, which shows the strength of the emission lines remaining relatively constant during low and intermediate states of continuum flux, but rising sharply during the strongest continuum flares. 

For continuum fluxes between 1.0$\times 10^{-14}$ and 2.8$\times 10^{-14}$ erg~s$^{-1}$~cm$^{-2}$, the emission line fluxes display a wide range of values that appear uncorrelated with the continuum. In particular, the emission line fluxes during the first half of orbit~2 ($0.15\leq\phi\leq0.22$) are very low, even fainter than the emission during the eclipse of orbit~1. This could indicate a dependence of the line emission on the binary orbital phase. However, given that the data covers only one orbit, a definitive conclusion is not currently possible.

\subsection{Continuum Color}

\begin{figure}[ht!]
\plotone{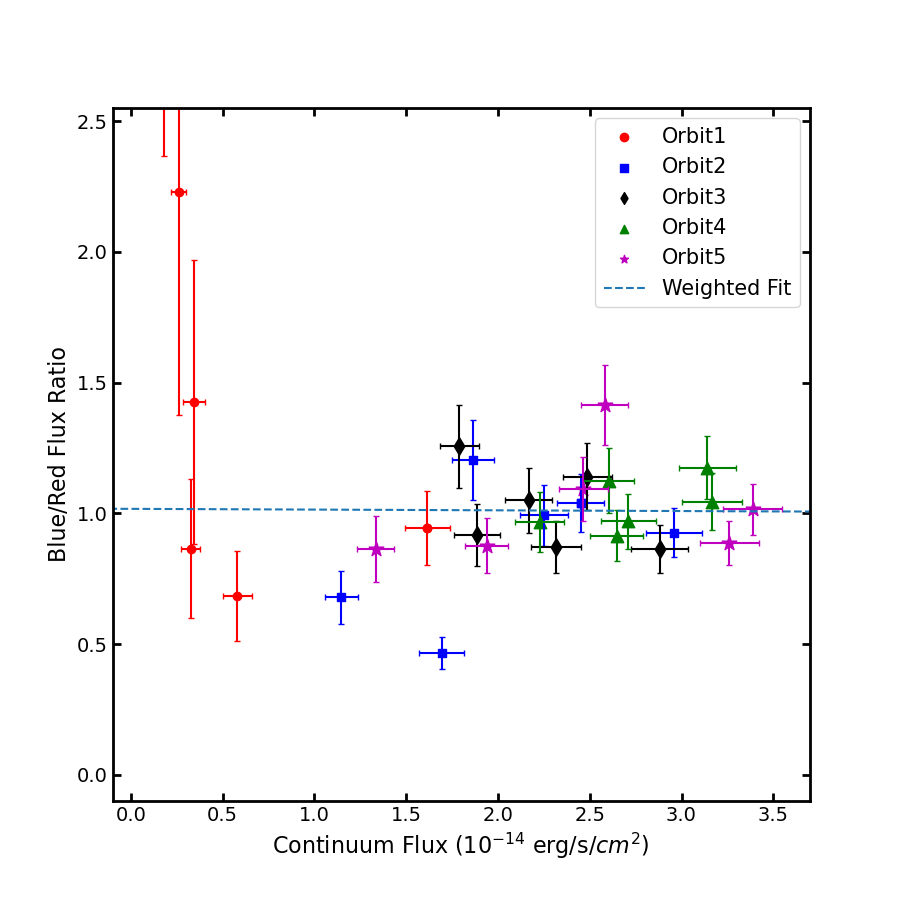}
\caption{ The continuum color calculated from the ratio of the flux in the three blue continuum bands to the flux in the three red continuum bands against the total flux from all six continuum bands. The dashed blue line is a weighted linear fit showing no significant relation between the color ratio and the continuum flux. Points corresponding to orbit~1 do not contribute strongly to this fit due to the low continuum flux during the eclipse.
\label{fig:BlueRed_Continuum}
}
\end{figure}

The continuum fluxes measured from the 30 spectral sub-exposures were used to investigate the color variations as a function of total continuum flux. Typical fractional uncertainties are 0.06 for the blue continuum and 0.09 for the red continuum. However, due to the low number of counts during the eclipse in orbit~1, these fractional uncertainties increase dramatically, especially in the red, leading to poorly constrained values for the final ratios. The blue/red flux ratio vs. continuum flux for all of the time samples is shown in \autoref{fig:BlueRed_Continuum}. 

A weighted fit was applied to the color vs. continuum flux points where the weights were calculated as the inverse of the quadrature sum of the fractional uncertainties in the blue and red continuum flux bands. As a result, the eclipse in orbit~1 does not contribute strongly to the fit. Overall, the FUV continuum color shows no significant slope during flares. The color may become slightly redder when the system is faint while out of eclipse. The fluxes during eclipse are too low to allow a confident estimate of the FUV color.

\subsection{UV Line Profile Variations Over a Binary Orbit}

\begin{figure}
\plotone{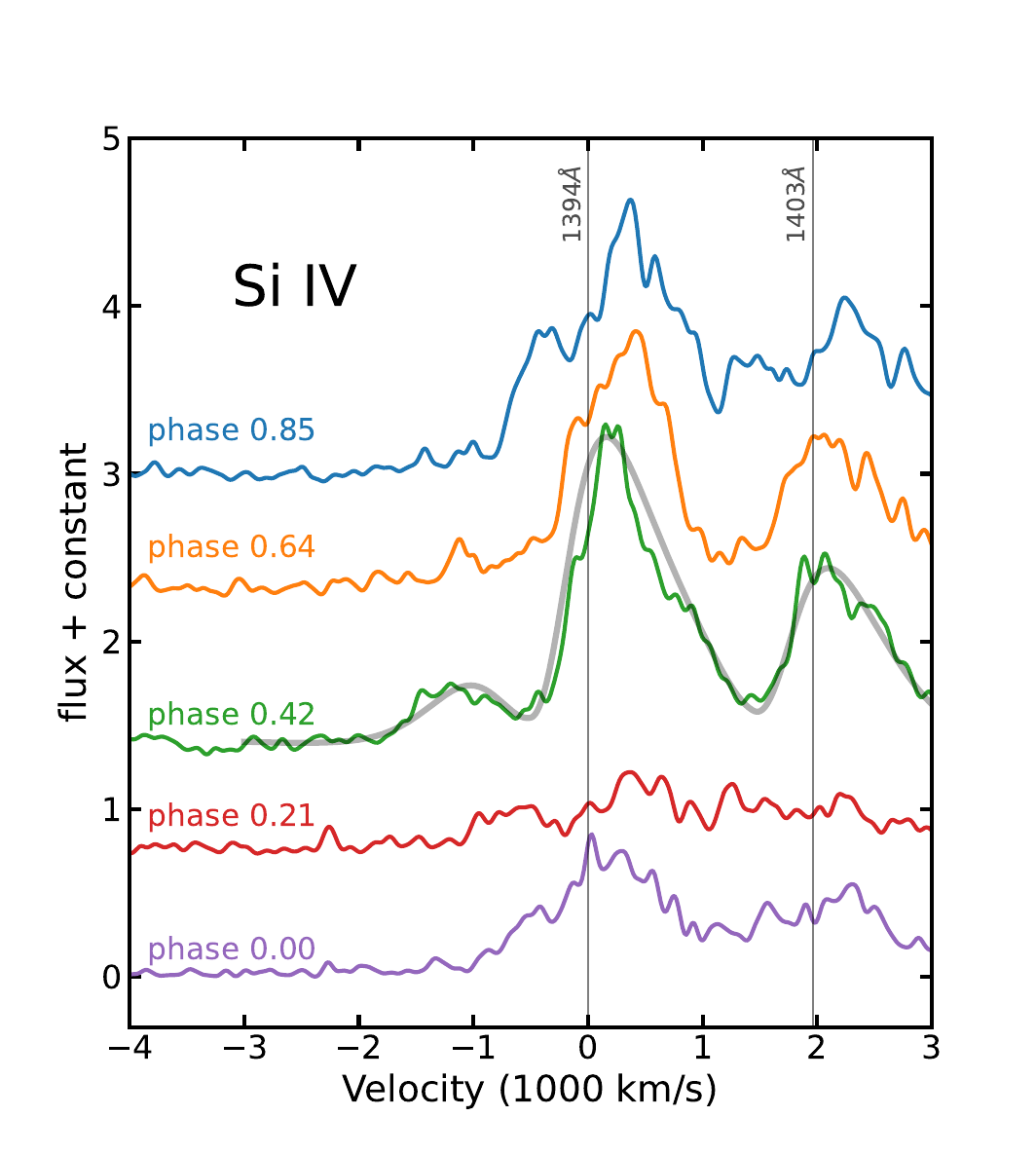}
\caption{The Si~IV emission doublet averaged over each of the five $HST$ orbits plotted versus the binary center of mass velocity. A constant has been added to vertically shift each spectrum for clarity. The average binary orbital phase is given on the left of each spectrum. The peak of the \ion{Si}{4} emission is redshifted relative to zero velocity over the entire binary orbit. The gray band over-plotted on orbit~3 shows a symmetric emission feature absorbed by a foreground outflow and applied to both components of the doublet.
\label{fig:si4}}
\end{figure}

The peaks of the UV resonance lines in the J0240 spectrum are red-shifted relative to the systemic velocity measured from the secondary star's orbit \citep{Thorstensen2020, Garnavich2021}. This is best seen in the Si~IV 1394/1403~\AA\ doublet because it is bright and has a large separation in its two components. The average Si~IV profile for each $HST$ orbit is shown in \autoref{fig:si4}. The asymmetry in the line profiles are particularly apparent in $HST$ orbits 3 and 4 covering binary orbital phases 0.4 to 0.7. The emission sharply drops off at $-300$ to $-400$~\kms , while by orbital phase 0.85, the Si~IV emission extends out to $-$800~\kms . This velocity range is similar to the narrow H$\alpha$ absorption detected over the same orbital phases in optical spectroscopy \citep{Garnavich2021}. For Si~IV, the absorption appears very broad and suggests a high-velocity outflow with a significant optical depth.

The \citet{Wynn1997} propeller model predicts that the highest blue-shifted radial velocities for the accelerated gas to be visible between orbital phases 0.4 to 0.6 depending on the assumed drag coefficient. Indeed, the \ion{Si}{4} spectrum at phase 0.42 shows an emission component peaking at $-1200$~\kms\ and extending to $-1600$~\kms . The asymmetric profile can be explained as an optically thick outflow in the resonance lines, with much of the blue-shifted emission absorbed by the earlier ejected gas. This P-Cygni-like situation is similar to the two-wind scenario considered by \citet{Fernandes1999}, where a continuum source is replaced by emission from an inner, fast wind.

We approximated the absorption profile using a function based on the plane-parallel expanding atmosphere calculation by \citet{Castor1979} and \citet{Noerdlinger1974}. A symmetric emission profile with a half-width of 1000~\kms\ was placed behind the absorbing atmosphere with an optical depth of $\tau >> 1$. The resulting line profile is a good match to the observed Si~IV line from $HST$ orbit~3 ($\phi=0.42$). Given the asymmetry of the propeller model, it is not clear how to extend this rough calculation to spectra at other orbital phases.

The high inclination of J0240 means that optical depth plays an important role in the resonance line profiles and their emergent flux. In addition to the narrow absorption seen in optical spectra, a broad, wind-like absorption may also shape the Balmer lines over the second half of the binary orbit.

\begin{figure}[ht!]
\plotone{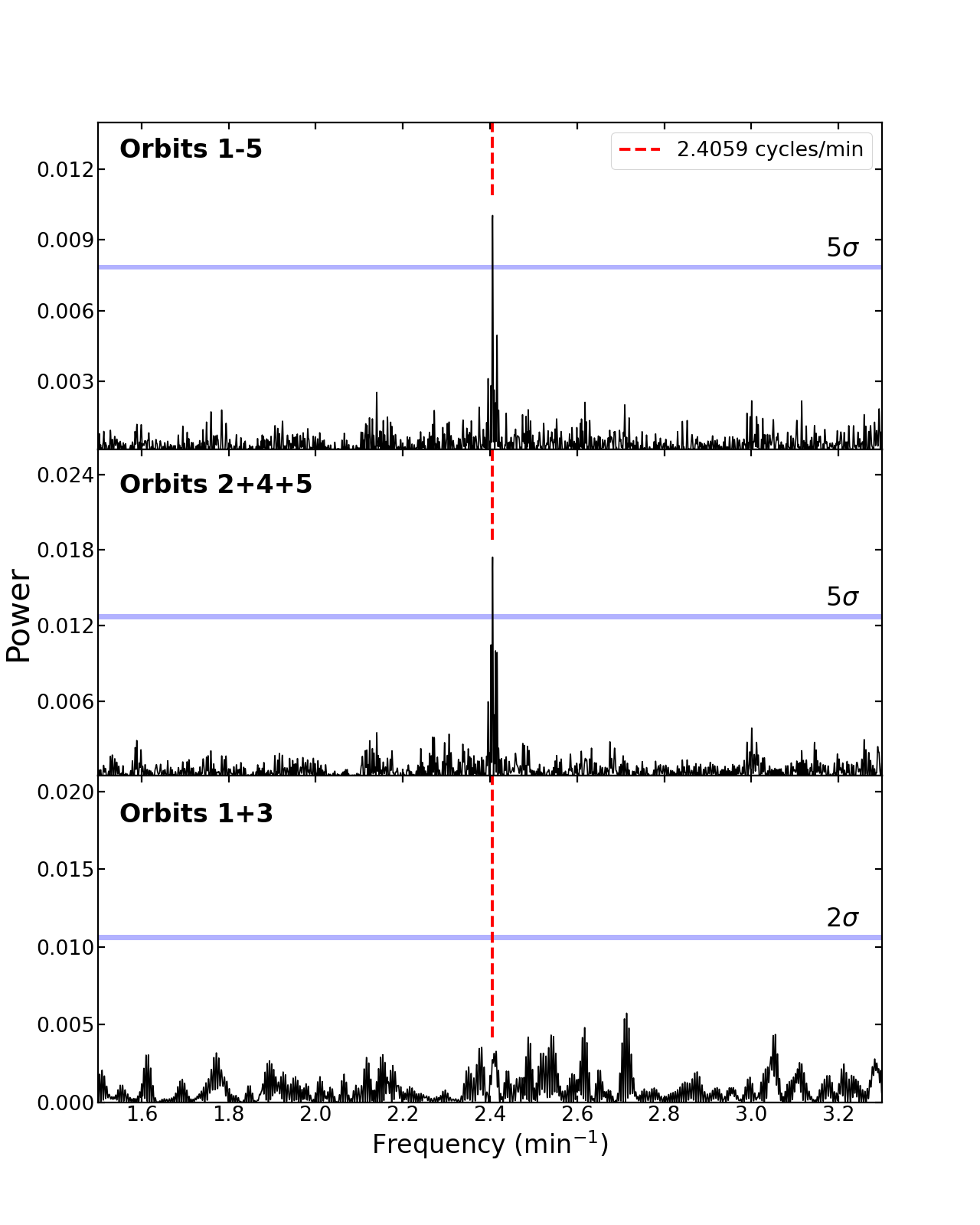}
\caption{Lomb-Scargle periodograms of the continuum (red plus blue bands) for $HST$ orbits 1-5 (top panel), the combination of orbits 2, 4, and 5 ($\phi=0.21,~0.64,~0.85$; middle panel), and the combination of orbits 1 and 3 ($\phi=0.00,~0.42$; bottom panel). The 24.939~s (2.4059~min$^{-1}$) WD spin pulse is more pronounced in orbits 2, 4, and 5, and has no significant signal in orbits 1 and 3. The spin pulse is not detected in the emission lines. The blue horizontal lines indicate a significance threshold for each power spectrum.
\label{fig:LS_Continuum}}
\end{figure}

\subsection{Detection of WD Spin Modulation}

Light curves were constructed by sorting the measured count rate from \textit{corrtag} files into 2~s wide bins. `Full spectrum' light curves were generated by integrating counts over the effective wavelength coverage of the detector (1150-2000~{\AA}), but with wavelength masks for telluric Ly-$\alpha$ and \ion{O}{1} emission. `Continuum' light curves were constructed from the six continuum bands detailed in Section \ref{Emission Line Strengths}. `Emission line' light curves were constructed from the combined regions of spectrum containing the \ion{C}{3}, \ion{N}{5}, \ion{C}{2}, \ion{Si}{4}, \ion{N}{4}, \ion{C}{4}, \ion{He}{2}, and \ion{Si}{3} emission lines.

The light curves were pre-whitened by fitting a third order polynomial to each orbit and subtracting it from the data. This was performed on the full spectrum as well as the emission line, and continuum light curves orbit by orbit. We then constructed Lomb-Scargle periodograms \citep[LSPs;][]{lomb76,scargle82} using the LombScargle function in astropy's timeseries package \citep{Astropy} for the light curves of each orbit and the light curves of various combinations of orbits. 

The full spectrum LSPs constructed from the combination of all five orbits revealed a strong periodic signal near 2.4 cycles~min$^{-1}$, which is interpreted as being due to the spin of the WD as found in ground-based photometry by \citet{Pelisoli2022}. This periodic signal is strongly detected in the continuum-only periodograms, but absent in the emission line LSPs. The continuum LSPs for the combination of all five orbits, a combination of orbits 2, 4, and 5 ($\phi=0.21,~0.64,~0.85$), and a combination of orbits 1 and 3 ($\phi=0.00,~0.42$) are shown in \autoref{fig:LS_Continuum}. 

\begin{figure}[ht!]
\plotone{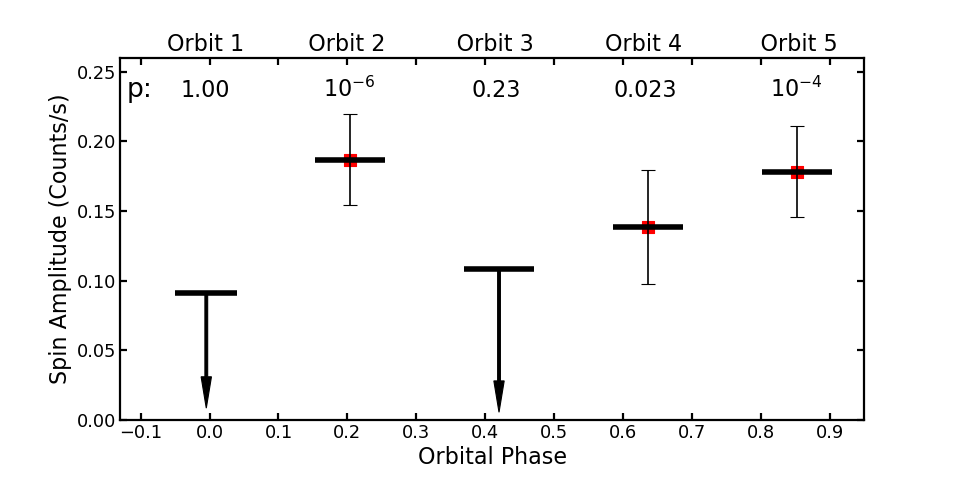}
\caption{Amplitude of the peak in the continuum periodogram that is closest to 2.4059 cycles~min$^{-1}$. The p-value for each orbit indicates the probability that the signal is due to chance. Orbit~1 does not have a measurable peak, instead the upper limit of a 1$\sigma$ detection limit is shown. Orbit~3 has a measurable peak just above a 1$\sigma$ detection limit, but its p-value means its significance is unlikely.
\label{fig:ContinuumCounts_vs_Phase}}
\end{figure}

The amplitude of the continuum spin pulse was estimated for each orbit by adding a series of sinusoidal modulations to the light curves and varying their counts/s amplitude until they matched the power of the spin pulse in the LSPs. The amplitude of the spin pulse in the continuum bands is shown in \autoref{fig:ContinuumCounts_vs_Phase} as a function of orbit number and orbital phase. To estimate the significance of peaks in the periodograms of each orbit, we shuffled the 2~s count rate bins, and this was repeated 10$^5$ times. The maximum power spectrum peak between 2.1 and 2.7 cycles~min$^{-1}$ was then recorded for each shuffled light curve. From this we calculated a p-value for the spin signal of each orbit, which quantifies the likelihood that observed peak could occur at random. Orbit~1, which coincides with the eclipse of the system, shows no significant peaks in its power spectrum. The spin signals in orbits 2, 4, and~5 have p-values less than 0.05, implying that they are highly likely to be significant. 

Orbit~3 shows a weak signal near 2.4 cycles~min$^{-1}$ in its continuum periodogram, but the strength is not significantly above a 1$\sigma$ lower limit of detectability. There is no such peak in the full power spectrum (1150-2000~\AA) of orbit~3. We consider the peak near 2.4 cycles~min$^{-1}$ in the continuum-only power spectrum of orbit~3 to be a non-detection of the spin pulse. 

From the continuum-only LSP constructed with the light curves of orbits 2 through 5 ($0.15\leq\phi\leq 0.90$), we measured the frequency of the spin signal to be 2.4059~$\pm$~0.0006 cycles~min$^{-1}$. This corresponds to a period of 24.939~$\pm$~0.006~s, consistent with the periodicity discovered by \citet{Pelisoli2022}. Orbit~1 was excluded from this analysis as it corresponds to the time of eclipse. The frequency of the signal shows a small variation between the LSPs constructed from the continuum-only and the full spectrum light curves, and this was used to estimate the uncertainty on the adopted modulation frequency. 

We expect that the spin pulse arrival times from J0240 would vary over the binary orbit, as this was observed in AE~Aqr by
\citet{Eracleous1994}. The FUV continuum flux in J0240 is too weak to detect individual pulses, so the varying pulse timing can not be directly measured. We modeled the impact of the pulse timing variations on the LSP constructed over the entire orbit and found that it lowered the amplitude of the spin peak, but had little impact on the recovered frequency.

\begin{figure*}[ht!]
\plotone{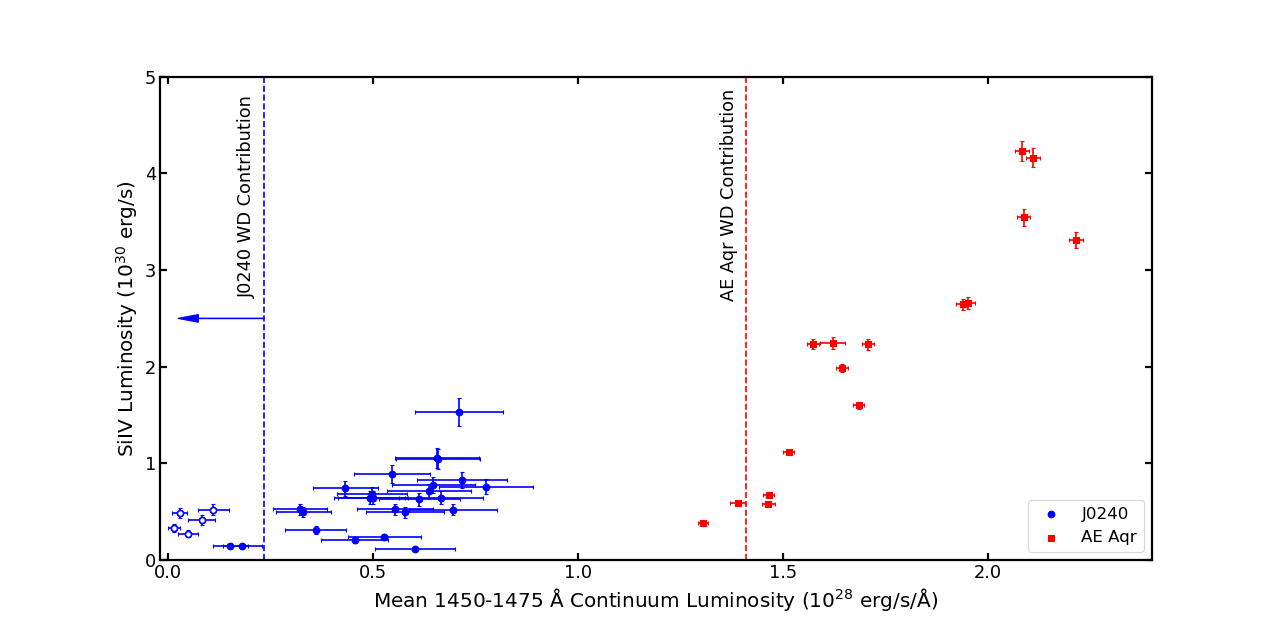}
\caption{ The relation of the \ion{Si}{4} and continuum luminosities for J0240 and AE Aqr. Points representing J0240 come from the 30 sub-exposures created from the five $HST$ orbits. Open circles for J0240 indicate the eclipse during orbit 1. Points representing AE Aqr are taken from the 16 total $HST$ orbits in 1992 and 1993. Vertical dashed lines indicate the estimated contribution of each system's WD to the total continuum.
\label{fig:Luminosity_SiIV}}
\end{figure*}

\subsection{J0240 and AE~Aqr FUV Luminosity Comparisons} \label{Luminosities}

Here, we directly compare the behavior of the FUV emission lines and continuum of J0240 and AE Aqr. We downloaded publicly available $HST$-FOS (Faint Object Spectrograph) archival observations of AE~Aqr taken in 1992 (Proposal ID: GO-3600, PI: K. Horne) and 1993 (Proposal ID: GO-5255, PI: K. Horne) reduced through the FOS calibration pipeline. The data from both AE Aqr FOS observations can be found in MAST, DOI:\dataset[10.17909/ma0a-qb57]{http://dx.doi.org/10.17909/ma0a-qb57}. 

FOS wavelength calibrations were known to have errors reaching more than 1000~\kms\ in the low dispersion gratings \footnote{\href{https://www.stsci.edu/documents/dhb/pdf/c07_foserrors.pdf}{STScI FOS Data Handbook Chapter 32: Error Sources}}, and there was a clear offset between the emission lines in the AE~Aqr and J0240 spectra. To correct the wavelength solution error, we shifted the wavelengths of the FOS spectra by a factor of 1.006 to match the \ion{Si}{4} peaks in the AE~Aqr FOS and J0240 COS spectra. We defined a new continuum band between 1450-1475~{\AA}, where the 2022 J0240 COS and the 1992 and 1993 AE~Aqr FOS spectra show no obvious spectral features. Emission line and continuum fluxes for AE~Aqr were calculated in a similar manner to J0240. 

We used $Gaia$ DR3 parallaxes to calculate the distance to J0240 (617.99~$\pm$~27.06~pc) and AE Aqr (91.65~$\pm$~0.15~pc) \citep{Gaia2016,Gaia2023}. We also corrected for extinction effects using the 3D Bayestar19 dustmap \citep{Green2019} in the astropy dustmaps package. The sight line to J0240 has a reddening of E(B-V) = 0.02 at a distance of 618~pc. The sight line to AE Aqr has negligible extinction at a distance of 92~pc. Wavelength-specific extinctions are calculated for each emission line using a linear interpolation of the values presented in \citet{Fitzpatrick2019}. Uncertainties in the extinction flux corrections were estimated by calculating the difference in the corrections on the scale of a half-width of the extinction coefficient wavelength bins. Average reddening flux corrections for J0240 are around 11.5\% across the full COS spectrum. The $Gaia$ distances and reddening corrected fluxes are then used to calculate luminosities for emission lines and the 1450-1475~{\AA} continuum band.

We defined high and low states of continuum flux for J0240 as the sub-exposures with the 8 highest and 8 lowest ``total" (red + blue) continuum flux measurements. This excludes the sub-exposures in orbit~1 due to the presence of the eclipse. The high-state sub-exposures are then averaged to create a mean high-state luminosity, the same is done for the low states to create a mean low-state luminosity. 

For AE~Aqr, we define a mean high-state luminosity by averaging two of the brightest orbits of the 16 total observed over the two FOS programs. The low-state average is found by averaging the two lowest luminosity orbits. The emission line and continuum luminosities for J0240 and AE~Aqr are compared in \autoref{tab:LuminosityComparison}.

During flares, the continuum flux and emission line strengths are well correlated in AE~Aqr \citep{Eracleous+Horne1996}. A comparison between the variation in the luminosity of the \ion{Si}{4} emission as a function of the continuum luminosities of the two systems is presented in \autoref{fig:Luminosity_SiIV}.
It is clear that the FUV luminosity of J0240 is significantly less than that of AE~Aqr. Further, the flaring amplitudes in the continua and emission lines are greater in AE~Aqr.

\autoref{fig:Luminosity_SiIV} also indicates our estimated continuum contributions from the WDs. For J0240, we estimated an upper limit of the WD flux in the 1450-1475~{\AA} continuum band to be the mean flux density over those wavelengths during the lowest flux portion of orbit 2 ($0.16\leq\phi\leq0.21$). The white dwarf in AE~Aqr is estimated to contribute 1~mJy in this continuum band, which is roughly the mean flux density of the WD and its periodic spin signal in Figure~14 of \citet{Eracleous1994}. We find that the WD in J0240 is about six times less luminous than its counterpart in AE~Aqr.

\subsection{J0240 and AE~Aqr Optical Comparisons} \label{optical}

We can compare the optical flaring amplitudes of J0240 and AE~Aqr using the transient surveys ASAS-SN \citep{Shappee2014,Hart2023} and ATLAS \citep{Tonry2018,Heinze2018}. These surveys sample the light curves with a cadence of a few days, so variability of flares must be analyzed using statistical methods. Flares from AE~Aqr are saturated in the ATLAS data, so the $g$-band photometry from ASAS-SN is used for this very bright star. J0240 is too faint for ASAS-SN monitoring, so the $c$-band photometry from ATLAS was used to analyze the flaring in J0240. 

The ATLAS $c$-band is very broad and has an effective wavelength between the $g$ and $V$-bands. Using spectra taken during a flare event and published in \citet{Garnavich2021}, we integrated the fluxes over both bandpasses and estimate that J0240 brightness measurements in the ATLAS $c$-band would be 0.25~mag brighter than if it were observed in the ASAS-SN $g$-band used for AE~Aqr.

\begin{figure}[ht!]
\plotone{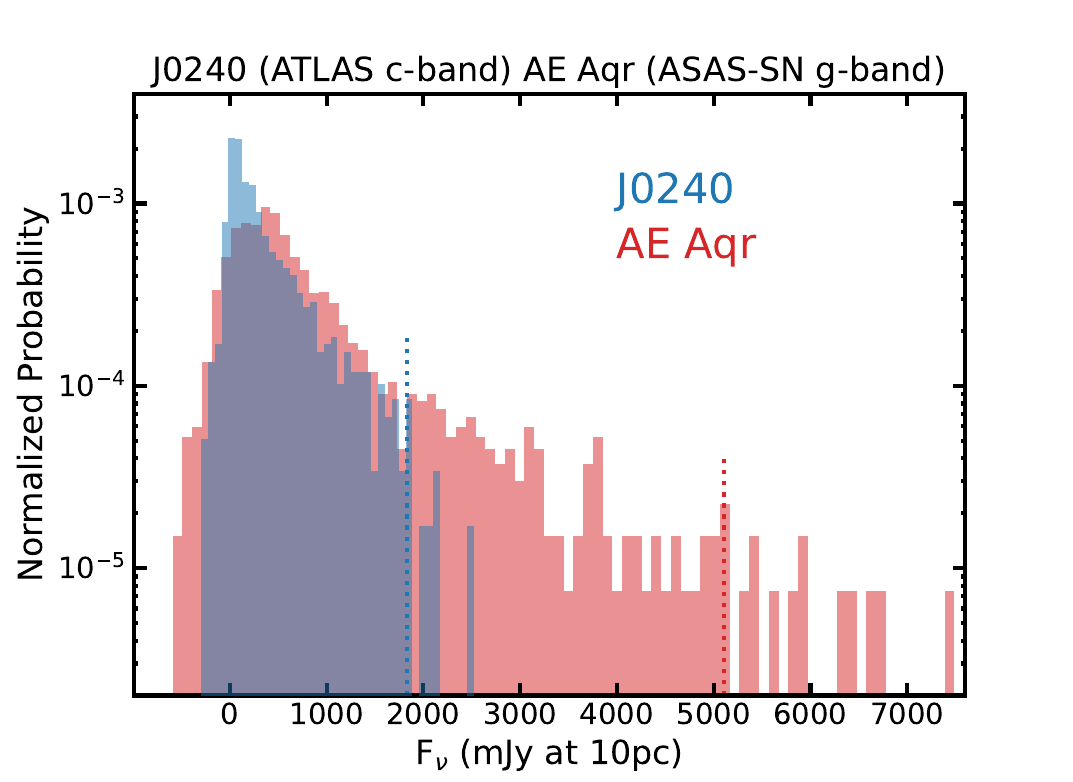}
\caption{The distribution of optical luminosities for J0240 (blue) and AE~Aqr (red) after the subtraction of their secondary star contributions and scaled to a common distance of 10~pc. The J0240 photometry was from the ATLAS $c$-band while AE~Aqr was measured from the ASAS-SN $g$-band. 99\%\ of the flux measurements fall below the dotted vertical lines, showing that the brightest optical flares from AE~Aqr are 2.8 times more luminous than from J0240.
\label{fig:optical}}
\end{figure}

We phased the 838 $c$-band ATLAS observations of J0240 on its orbital period and subtracted the ellipsoidal modulation of the secondary star. We deleted 92 observations taken during the eclipse by removing measurements with orbital phases of $\pm 0.05$. We scaled the remaining fluxes to a distance of 10~pc and show the distribution of the flare luminosities in 
\autoref{fig:optical}. We found that 99\%\ of the observations are fainter than $L_{max}=1850$~mJy if J0240 were 10~pc away, and that is our estimate for the maximum luminosity of the flares. 

We repeated this analysis for the 5323 $g$-band measurements of AE~Aqr from ASAS-SN. Note that ASAS-SN takes multiple exposures when it visits a field, so we averaged flux measurements that were taken within a span of a few minutes, leaving 1324 independent ASAS-SN brightness estimates. The distribution of measured fluxes for AE~Aqr, scaled to a distance of 10~pc, are displayed in \autoref{fig:optical}. We found that 99\%\ of the measurements are fainter than $L_{max}=5111$~mJy, suggesting that flares in AE~Aqr reach 2.8 times greater optical luminosities than in J0240. Due to the differences in bandpasses used, we expect this ratio is a lower limit to the true ratio. The continuum flares in AE~Aqr peak at higher luminosities than in J0240 at both the FUV and optical wavelengths.

After normalizing the luminosity distributions by their respective $L_{max}$ values, the cumulative histograms imply a difference in the flare luminosity distributions between AE~Aqr and J0240. A Kolmogorov-Smirnov analysis of the distributions gives a KS statistic of $D=0.13$ and a probability of $10^{-7}$ that the two observed data sets were drawn from the same distribution. 

The normalized flare distributions are approximately power-laws where the probability, $P$, of an observed flare with luminosity $L$ is $P[L/L_{max}]\propto (L/L_{max})^{-\alpha}$. For AE~Aqr, the least-squares fit to the $L/L_{max}$ distribution is $\alpha= 5.22\pm 0.35$, which is significantly steeper than the fit for J0240 which yields a power-law index of $\alpha=3.85\pm 0.22$. While AE~Aqr has more luminous optical flares as shown by an $L_{max}$ that is $>2.8$ times larger than for J0240, the luminosity distribution of AE~Aqr falls off more rapidly than for J0240. Photometric samples of the these two propellers suggest that a random visit is more likely to find J0240 near its maximum luminosity compared with AE~Aqr. This could mean that continuum flares in J0240 decline more slowly than in AE~Aqr, or that flares from J0240, while weaker, occur more often than AE~Aqr.

\subsection{Simultaneous UV and Optical Photometry}

\begin{figure}
\plotone{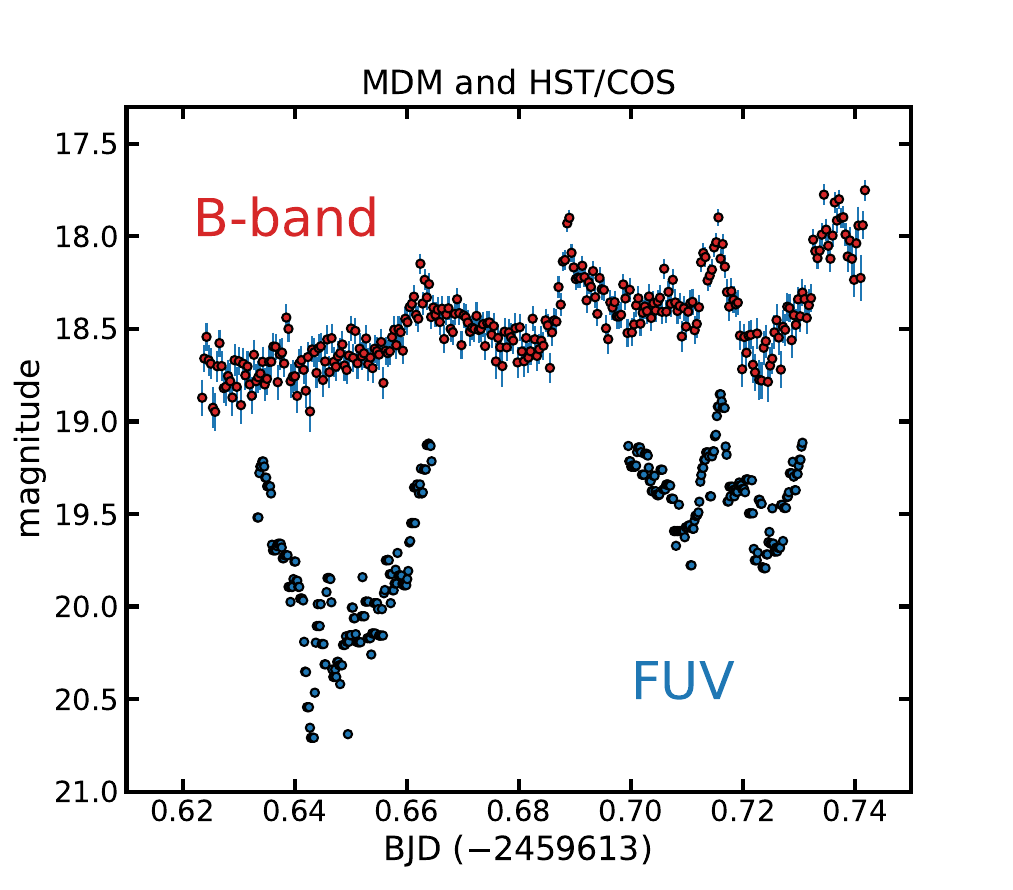}
\caption{The MDM B-band light curve (red points) obtained simultaneously with the $HST$ orbits~2 and 3 ($0.15\leq\phi\leq 0.47$). For a direct comparison, the total counts from the FUV spectra have been divided into 15~s bins and converted to magnitudes with an arbitrary zero-point. Flaring in the optical light curve corresponds to peaks in the FUV data. \label{fig:mdm_hst}}
\end{figure}

Ground-based optical photometry from the MDM Observatory were obtained simultaneously with the $HST$ observations. The optical data spanned the binary orbital phases between 0.13 and 0.51, and overlapped with orbits 2 and 3 of the $HST$-COS observations. For comparison, we converted the flux of the full UV spectrum into magnitudes and set the zero-point so that the UV light curve is just below the calibrated $B$-band light curve as shown in Figure~\ref{fig:mdm_hst}. There is a slow brightening over the course of the observations in both the optical and FUV. The flare in the ground-based data corresponds with the higher amplitude flares in the FUV. The flares appear weaker in the optical due to the contribution of the secondary star. During the eclipse of the flaring region, the brightness of J0240 was estimated to be $B=19.0$ mag \citep{Garnavich2021}, and ellipsoidal variations mean the secondary is even brighter around quadrature \citep{Thorstensen2020}. Thus, the secondary adds significantly to the $B$-band light and dilutes the flare amplitudes when compared to the FUV where the secondary contribution is insignificant. 

We can directly compare the $B$-band and FUV amplitudes of the flare observed in the middle of orbit~3 ($\phi=0.42$). Averaged over the six continuum bands (Figure~\ref{fig:hst_spec}), the FUV flux increased by $4.8\times 10^{-17}$~\flam . The rise in the $B$-band corresponded to an increase of $1.5\times 10^{-16}$~\flam , or a factor of three larger in amplitude than in the FUV.

\begin{deluxetable*}{cccccccc}
    \tablecaption{J0240 \tablenotemark{{\footnotesize a}} and AE Aqr \tablenotemark{{\footnotesize b}} Low and High-state Luminosity Comparison}

    \tablenum{2}

    \tablehead{\colhead{ Ion } &\colhead{ J0240 Orbit 1 } & \colhead{J0240 Low} & \colhead{J0240 High} & \colhead{J0240 Ratio}\tablenotemark{{\footnotesize c}} & \colhead{AE Aqr Low} & \colhead{AE Aqr High} & \colhead{AE Aqr Ratio}\tablenotemark{{\footnotesize c}}\\ 
    \colhead{  } & \colhead{({10$^{29}$ erg/s})} & \colhead{(10$^{29}$ erg/s)} & \colhead{(10$^{29}$ erg/s)} & \colhead{  } & \colhead{(10$^{29}$ erg/s)} & \colhead{ (10$^{29}$ erg/s) } & \colhead{  }} 

    \startdata
        Continuum \tablenotemark{{\footnotesize d}} & 0.27~$\pm$~0.04 & 0.94~$\pm$~0.10 & 1.57~$\pm$~0.16 & 1.67~$\pm$~0.13 &  3.46~$\pm$~0.03 &  5.24~$\pm$~0.04 & 1.51~$\pm$~0.02 \\
        C III    & 0.43~$\pm$~0.07 & 0.39~$\pm$~0.05 & 0.69~$\pm$~0.09 & 1.78~$\pm$~0.24 &                  &                  &  \\
        N V      & 3.97~$\pm$~0.37 & 3.62~$\pm$~0.33 & 6.51~$\pm$~0.58 & 1.80~$\pm$~0.06 &  6.53~$\pm$~0.57 &  19.8~$\pm$~1.4  & 3.03~$\pm$~0.34 \\
        C II     & 0.43~$\pm$~0.06 & 0.48~$\pm$~0.06 & 1.78~$\pm$~0.24 & 1.97~$\pm$~0.20 &                  &                  &  \\
        Si IV    & 3.87~$\pm$~0.36 & 4.39~$\pm$~0.40 & 8.79~$\pm$~0.79 & 2.00~$\pm$~0.06 &  4.98~$\pm$~0.21 &  42.2~$\pm$~0.7  & 8.47~$\pm$~0.39 \\
        N IV     & 0.37~$\pm$~0.06 & 0.52~$\pm$~0.07 & 0.67~$\pm$~0.08 & 1.29~$\pm$~0.18 &  0.35~$\pm$~0.07 &  0.95~$\pm$~0.14 & 2.71~$\pm$~0.68 \\
        C IV     & 2.16~$\pm$~0.23 & 2.34~$\pm$~0.23 & 5.47~$\pm$~0.51 & 2.34~$\pm$~0.13 &  0.20~$\pm$~0.07 &  2.36~$\pm$~0.36 & 11.8~$\pm$~4.5 \\ 
        He II    & 0.36~$\pm$~0.08 & 0.42~$\pm$~0.08 & 1.00~$\pm$~0.14 & 2.36~$\pm$~0.50 &  0.80~$\pm$~0.18 &  4.77~$\pm$~0.36 & 5.9~$\pm$~1.4\\ 
        Si III   & 1.06~$\pm$~0.27 & 1.55~$\pm$~0.23 & 2.99~$\pm$~0.38 & 1.94~$\pm$~0.30 &  0.80~$\pm$~0.14 &  4.02~$\pm$~0.36 & 5.00~$\pm$~0.99 \\ 
    \enddata
    \label{tab:LuminosityComparison}
    \tablenotetext{a}{ Luminosities are calculated using flux measurements for each line combined with the distance to J0240 derived from the parallax measurements in $Gaia$ DR3, 618.0~$\pm$~27.1~pc. Luminosities have been corrected for extinction.}
    \tablenotetext{b}{Luminosities are calculated using the publicly available 1992 and 1993 $HST$-FOS pipeline calibrated fluxes and the distance to AE Aqr reported in $Gaia$ DR3, 91.65~$\pm$~0.15~pc. The low-state luminosities are an average of orbit~8 in 1992, and orbit~8 in 1993. The high-state luminosities are an average of orbit~3 in 1992, and orbit~2 in 1993. Extinction is negligible. }
    \tablenotetext{c}{ The ratio of high to low-state luminosity. Uncertainties in the ratio are calculated using the flux uncertainties. }
    \tablenotetext{d}{ The 1450-1475~{\AA} continuum band used to compare J0240 and AE Aqr, integrated over its width.}
\end{deluxetable*}

\subsection{WD Temperature} \label{White Dwarf Temperature Limits}

The WD may contribute a significant fraction of the FUV flux in J0240 since CV propellers are not expected to develop accretion disks, and their cool secondaries are not UV bright. The J0240 WD is not clearly detected in the FUV continuum, but we can place an upper limit on the temperature of the WD. As we are measuring the flux over many spin cycles of the WD, we can only constrain its average surface temperature. 

We utilize the model WD spectra from \citet{Koester2010} with modifications provided by \citet{Tremblay2009}\footnote{The Koester WD model spectra are available at \href{http://svo2.cab.inta-csic.es/theory/newov2/index.php?models=koester2}{Koester Models}}. The WD models predict surface flux density for a range of temperatures at a given surface gravity, $log(g)$. WD masses were calculated from the $log(g)$ values provided in these models using $log(g) = log(GM/R^{2})$, with the radius constrained by the mass-radius relation for a cool WD \citep{Nauenberg1972}\footnote{Radii calculated with this relation are strictly lower limits, as the radius of a WD increases with temperature at a given mass.}. The model flux density was scaled to the $Gaia$ distance, $D=617.99\pm 27.06$~pc, of J0240 using the relation $(R/D)^{2}$. J0240's WD rotation rate is approaching the breakup velocity, which puts a 0.7 M$_{\odot}$ lower limit on the mass of the WD \citep{Pelisoli2022}. As a result, we do not consider masses below this limit in our analysis. 

The FUV flux from the WD must compete with the flaring activity clearly present in the light curve. Therefore, the best time to search for the WD contribution is during the faintest quiescent periods. The continuum flux in J0240 is lowest over orbit~1 ($\phi=0.00$), but this is due to the secondary star passing in front of the flaring region and very likely blocking our view of the WD. Of the remaining data, the middle portion of orbit~2  ($0.16\leq\phi\leq0.21$) displays the lowest continuum flux. We created a spectrum from this faint section of orbit~2, and used it to estimate limits on the WD temperature. We generated a ``continuum spectrum" by masking out the FUV emission lines and smoothing the spectrum by performing a polynomial fit. We then compared the flux density at each wavelength between 1255-2000~\AA\ in the continuum spectrum of orbit~2 to the flux densities over the same wavelengths for each mass-temperature combination in the \citet{Koester2010} model WD spectra. Additional values were interpolated for mass-temperature combinations between those provided in the Koester models. Finally, we recorded the flux density ratio between each model and the continuum spectrum.

\begin{figure*}[ht!]
\plotone{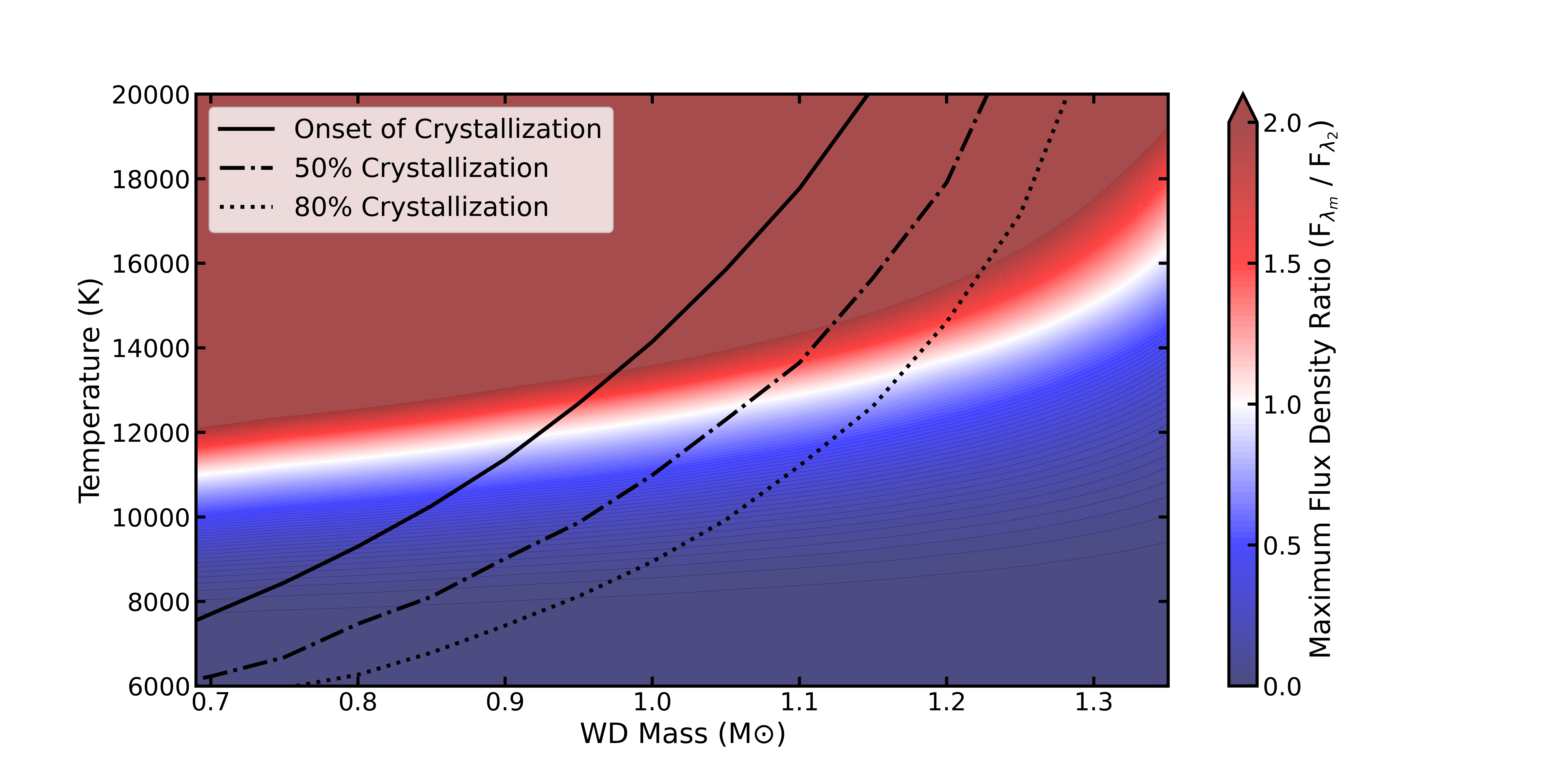}
\caption{The maximum ratio of the flux density of the WD model (F$_{\lambda_m}$) to the flux density of the continuum spectrum in the second $HST$ orbit (F$_{\lambda_2}$). A ratio of 1 indicates that the model flux is, at maximum, equal to the observed flux at some wavelength between 1255-2000~{\AA}, and can be interpreted as an upper limit of the mean WD temperature at a given mass-temperature combination. Mass-temperature combinations with a ratio above a value of 1 are ruled out as possibility. Combinations with ratios below a value of 1 are not ruled out since the models do not account for any source of continuum other than what is provided by the WD. Black lines showing the levels of crystallization of a CO WD illustrate where the magnetic dynamo scenario can operate.
\label{fig:Temperature_Mass}}
\end{figure*}

The mean temperature of the WD was constrained by this flux density ratio by assuming that a model spectrum can not be brighter than the observed continuum.  The model temperature was varied for a given mass, and if the model flux exceeds the observed flux at any wavelength (i.e. the maximum flux density ratio of model to observed is greater than 1), then that temperature was discounted as a possibility. If the ratio was less than or equal to 1 over the entire wavelength range, then model was deemed acceptable for the WD in J0240.  The ratio of the predicted to observed flux as a function of model mass and temperature are shown in Fig.~\ref{fig:Temperature_Mass}. The white band across the plot displays the upper limit of the average WD surface temperature as a function of the WD mass. For low masses, the average WD surface temperature is constrained to 11000~K or less. For high mass WDs, the average surface temperature is 15000~K or less.

\citet{Isern2017} proposed that rotating WDs with crystallized cores can generate a magnetic dynamo, and this idea is particularly interesting given the extremely rapid rotation rate for the WD in J0240. We applied the thin-atmosphere WD cooling models described in \citet{Bedard2020}\footnote{The WD cooling models are available at  \href{https://www.astro.umontreal.ca/~bergeron/CoolingModels/}{ Cooling Models}} to investigate the extent of WD crystallization that is allowable at a given mass-temperature combination. The calculated crystallization curves are overlaid on the temperature limits in \autoref{fig:Temperature_Mass}. These cooling models assume a carbon-oxygen (CO) core WD; the levels of crystallization are lower limits, as high-mass oxygen-neon core WDs (M~$\gtrsim$~1.1~M$_{\odot}$) will crystallize earlier at higher temperatures \citep{Blatman2024b}. If the WD in J0240 has a mass of 0.9~M$_{\odot}$ or greater, then it is likely to have begun crystallization. More massive WDs are expected to be in advanced stages of crystallization. There is, in fact, only a small wedge of temperatures and masses where crystallization is unlikely to have begun. 

\section{Discussion\label{sec:discussion}}

\subsection{Emission Line Ratios}

The COS spectrum of J0240, displayed in \autoref{fig:hst_spec}, reveals a significant \ion{C}{4} emission line as well as lower ionization carbon features. This is in contrast to the extremely weak carbon emission in the AE~Aqr propeller.

For a magnetic CV, J0240 displays a weak \ion{He}{2} emission line when compared with \ion{Si}{4}, and this is similar to the strength of \ion{He}{2} observed in AE~Aqr \citep{Eracleous+Horne1996}. The optical 4686~\AA\ emission of \ion{He}{2} is also unusually weak for a magnetic CV \citep{Thorstensen2020}, suggesting that these propellers have little or no direct accretion on to the magnetic poles of the WD as seen in polars and IPs.

\begin{figure}[ht!]
\plotone{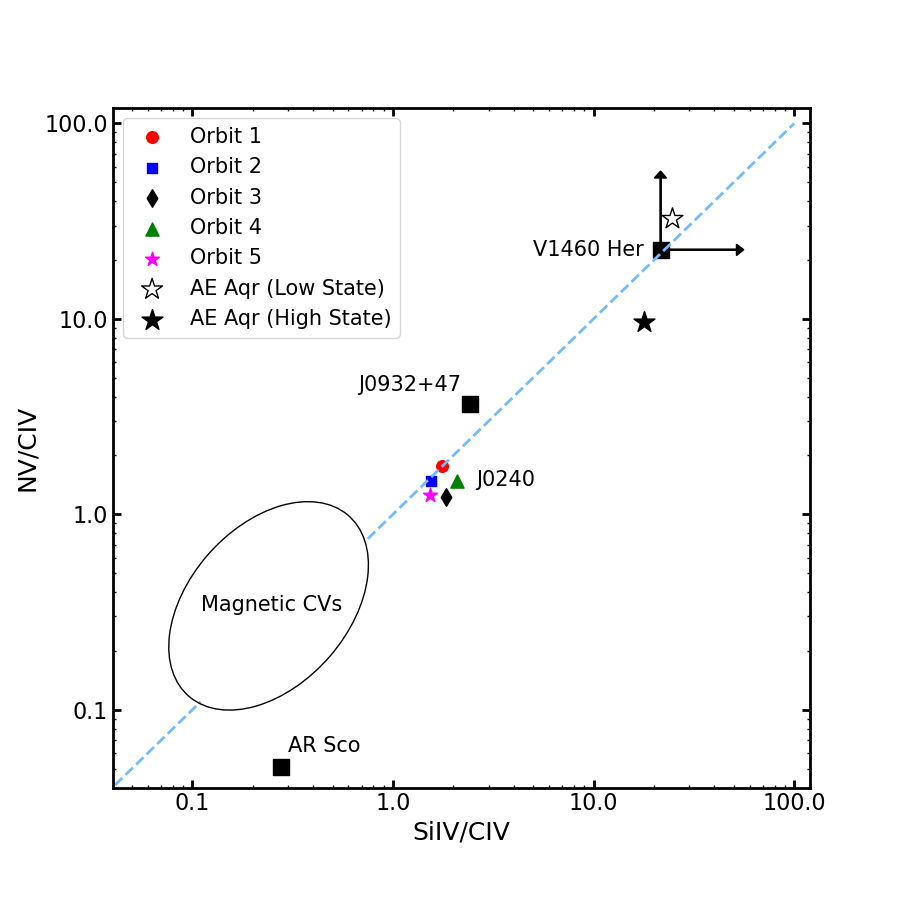}
\caption{\ion{N}{5}/\ion{C}{4} and \ion{Si}{4}/\ion{C}{4} emission line ratios for orbits 1-5 of the $HST$-COS observation of J0240, along with other WD binaries that exhibit high-frequency periodicities and have anomalous emission line ratios. ``Magnetic CVs'' are defined as in \citet{Mauche1997}. V1460 Her has a 38.9~s WD spin period, is non-magnetized, and contains a disk. The arrows indicate that the measured ratios are lower limits \citep{Ashley2020}. J0932+47 is a CV system with a disk that has an uncategorized 29.8~s periodic signal in its COS spectrum \citep{Tweddale2024}. AR Sco is a WD pulsar with a WD spin period of 117 s \citep{Gaibor2020}, the emission line ratios are calculated using publicly available COS data. The dashed line runs along \ion{N}{5}/\ion{C}{4} = \ion{Si}{4}/\ion{C}{4}. Error bars for the J0240 ratios are smaller than the size of the markers.
\label{fig:CIV_Ratios}}
\end{figure}

The average \ion{N}{5}/\ion{C}{4} ratio of 1.4~$\pm$~0.2 in J0240 is greater than what is typically seen in magnetic WDs. The newly measured \ion{N}{5}/\ion{C}{4} and \ion{Si}{4}/\ion{C}{4} ratios for J0240 are shown in \autoref{fig:CIV_Ratios}, as well as ratios for AE Aqr and the apparently non-magnetic CV V1460 Her \citep[spin period of 38.9~s;][]{Ashley2020}. Ratios for two other binaries with high-frequency periodicities: J0932+47 \citep{Tweddale2024} and the WD pulsar AR Sco \citep{Gaibor2020}, are also displayed. The \ion{C}{4} emission features in typical magnetic CVs tend to be stronger than the \ion{N}{5} and \ion{Si}{4} lines, with \ion{N}{5} and \ion{Si}{4} being approximately equal in strength \citep{Mauche1997}. However, CVs containing rapidly spinning WDs have shown weak \ion{C}{4} lines that result in anomalous \ion{N}{5}/\ion{C}{4} and \ion{Si}{4}/\ion{C}{4} ratios. For example, FUV spectra of AE Aqr show almost no \ion{C}{4} emission at all, leading to \ion{N}{5}/\ion{C}{4} flux ratios $>10$ \citep{Eracleous+Horne1996,Mauche1997}.

The anomalous \ion{N}{5}/\ion{C}{4} ratio in AE Aqr has been attributed to the system recently enduring a state of thermal-timescale mass transfer from an evolved secondary \citep{Schenker2002}. AE Aqr's long orbital period, relative to other CVs, suggests that the secondary would have had the opportunity to perform significant equilibrium CNO processing before filling its Roche Lobe. This would strongly suppress the relative carbon abundance in the innermost regions of the star, while simultaneously enhancing the relative nitrogen abundance. The rapid period of thermal-timescale mass transfer strips away the outermost layers of the secondary. This leads to an abnormally low carbon abundance while simultaneously spinning up the WD to its high rate of rotation \citep{Schenker2002}. This argument has also been applied to V1460 Her to explain its rapidly rotating WD and the absence of \ion{C}{4} from its spectrum \citep{Ashley2020}. 

We suspect that the presence of a sizable \ion{C}{4} emission line in the J0240 spectrum and its relative absence in AE Aqr is primarily related to the mass ratio of the binary components at the onset of mass transfer. The thermal-timescale mass transfer required to spin up the WD only occurs when the donor is more massive than the accretor, which in these cases are the secondary and WD respectively. A simple explanation would be that the ratio M$_{WD}$/M$_{2}$ would have been greater for J0240 than for AE Aqr at the onset of mass transfer. This would allow AE Aqr's WD to strip its secondary down to the more interior carbon-poor but nitrogen-rich layers. Similarly, the initial mass ratio can affect the time at which mass transfer begins, allowing for differences in the extent of CNO processing within the secondary. We plan on further work using the Modules for Experiments in Stellar Astrophysics (MESA) binary controls \citep{Paxton2015} to model the co-evolution of binary components that result in systems similar to J0240 and AE Aqr.

\subsection{Continuum and Emission Line Variability}

J0240 shows no significant variation in the blue to red continuum flux ratio in the FUV. This is in contrast to AE~Aqr, in which the continuum slope gets redder as the total continuum flux increases. \citet{Eracleous+Horne1996} attributed the color variation to free-bound Balmer recombination emission, which rises steeply into the NUV wavelengths. The lack of color changes in J0240 may be explained by the weak flares, and therefore smaller variations in the Balmer continuum. Further, the J0240 COS spectra were limited to wavelengths of $<2000$~\AA\, that are less sensitive to the Balmer continuum than in the NUV. 

AE~Aqr shows a clear linear correlation between increasing continuum flux and emission line strengths \citep{Eracleous+Horne1996}. This is well illustrated in \autoref{fig:Luminosity_SiIV} for \ion{Si}{4}. However, \autoref{fig:Flux_Continuum} shows that in J0240 the emission line strengths show little variation until the continuum reaches its highest amplitude. In fact, there is significant variation in emission line strength from orbit to orbit that does not correlate with the continuum flux. In particular, the emission lines over the first half of orbit~2 ($0.15\leq\phi\leq0.22$) are very weak, while its continuum is consistent with fluxes during orbits 3, 4, and 5 ($0.37\leq\phi\leq 0.90$). The emission line strengths over this portion of orbit~2 are even lower than those seen during the eclipse from orbit~1. 

The high orbital inclination of J0240 may make optical depth a significant factor in understanding the FUV emission line variability. Certainly the broad P-Cygni absorption seen in orbit~3 ($\phi=0.42$) suggests the existence of self-absorption in the resonance lines. But further data is needed to see how the emission varies with binary orbital phase.

The detection of emission lines during the eclipse, suggests that the source of some of the emission extends from the continuum flaring region and the WD. This is seen at optical wavelengths as H$\alpha$ emission is present during eclipse and has been seen to increase in brightness before the eclipse is complete \citep{Garnavich2021}. More spectra taken over eclipses could map the location of emission line regions.

\subsection{WD Spin Period}

We observe the WD spin period to be 24.939~$\pm$~0.006~s in the UV, consistent with the spin period found in the optical by \citet{Pelisoli2022}. The spin signal is easily detectable in the continuum portions of the UV spectrum, and not detectable within the emission lines. AE Aqr similarly does not show any periodicities within its emission line fluxes \citep{Eracleous1994}. Note that a WD spin, beat, and side-band modulations in IP systems are often detected in both the continuum and emission lines due to the interactions between the accretion curtains, partial disk, and stream.

We also find that the signal strength of the spin varies with orbital phase. The spin signal is strong in orbits~2, 4, and 5 ($\phi=0.21,~0.64,~0.85$). As expected, the spin pulse is not detectable during orbit~1, which coincides with the eclipse of the WD. Curiously, the spin pulse is not significant in orbit~3 ($\phi=0.42$), which closely coincides with superior conjunction. While the amplitude of the spin pulse may not change appreciably during the course of the binary orbit outside of the eclipse, the detectability of the spin pulse appears to be variable.

The continuum spin signal may appear stronger during flares. This is suggested when isolating both the flare in orbit~5 and the start of an apparent flare at the end of orbit~2. While orbit~3 does have what appears to be a small flare, we do not detect a spin signal when isolating this region of its light curve. 

The apparent increase in spin power during flaring matches what is seen in the optical power spectrum in Figure 3 of \citet{Pelisoli2022}, where there is excess power in the 2.4~min$^{-1}$ peak around phase 0.75 and just before eclipse, which both coincide with flares. In AE Aqr, the flares have been observed to have an effect on the optical spin signal \citep{Patterson1979}, and a minor but statistically significant effect on the amplitude of the UV spin pulse \citep{Eracleous1994}.

\subsection{WD Temperature and Crystallization}

We find that the average temperature of the WD in J0240 can be no more than $\sim$11,000 for a 0.7~M$_{\odot}$ WD and $\sim$15,000~K for a 1.3~M$_{\odot}$ WD. Applying the Koester models to the faintest FOS spectra from AE~Aqr provides an average temperature constraint that is similar to what was achieved with J0240. Assuming a 0.9 $M_{\odot}$ WD, we estimate an average temperature of 15,500~K for the WD in AE~Aqr. This is consistent with the temperature mapping done by \citet{Eracleous1994}. In contrast, our temperature upper limit at 0.9~$M_{\odot}$ for J0240 was about 11,700~K. The J0240 WD appears to be cooler than the one in AE~Aqr, and this may explain a large fraction of the luminosity difference shown in \autoref{fig:Luminosity_SiIV} when flaring is low. 

As the spectra do not resolve the WD spin in J0240, the constraints are on the average temperature over the WD surface. As in AE~Aqr, the presence of a spin modulation implies that the magnetic poles of the WD are heated producing a significant temperature gradient across the WD surface. Similarly, there is a significant portion of the surface of the WD that remains unheated. This temperature may better represent the cooler WD that would have been present prior to the onset of mass transfer. This means that each of the WDs in J0240 and AE Aqr could be in more advanced stages of crystallization than their average temperatures would suggest.

\subsection{Possible Evolutionary Scenarios}

A detached close WD-MS binary is created during a common envelope phase, resulting in a system that contains a MS component with a higher mass than the WD. It is possible that an initial magnetic field is generated during this phase \citep{Tout2008,Nordhaus2011}. WDs above 0.7~M$_{\odot}$ will cool to the point of the onset of crystallization between roughly 0.2 and 2~Gyrs after becoming a WD \citep{Bedard2020}, with the crystallization beginning earlier for more massive stars due to the increased pressure at their center. During this stage, as long as the WD has at least some rotation, it can generate a magnetic field at the onset of crystallization \citep{Fuentes2024}. This combination of conditions leading to crystallization and rotation-driven dynamo may explain the strong magnetic fields observed in some CVs \citep{Schreiber2021}. Based on the temperature constraints and the WD cooling models, at least some level of crystallization is likely for the WD in J0240. 

Thermal timescale mass transfer begins, and the WD is spun up on the order of 10$^{-7}$~M$_{\odot}$/yr \citep{Schenker2002}. A preexisting magnetic field from a slowly rotating WD can be heavily amplified during the spin-up phase \citep{Ginzburg2022}. There may be an equilibrium point where the WD can no longer be spun up during the thermal timescale mass transfer as the spin is fast enough and magnetic field strong enough that the material from the secondary will no longer be able to accrete onto the WD. Eventually, the mass transfer slows and the propeller is able to operate efficiently. As energy is expended in the ejection of gas, the WD spins down to the point where accretion can again begin, resulting in an intermediate polar system. 

Another consideration is that if the magnetic field is generated at the onset of crystallization, it will take roughly a Gyr for it to diffuse to the surface of the WD \citep{Ginzburg2022,Blatman2024a}. It is not clear if the strengthening of the magnetic field during accretion spin-up leads to a change in the timescale of magnetic field diffusion. There may be a scenario where after the spin-up phase the system must wait for the magnetic field to breakout from the surface, and the propeller mode immediately turns on once the strong magnetic field surfaces. The combined timescales of the crystallization of the WD and the diffusion of magnetic fields (roughly a Gyr each) could lead to a significant ``waiting time" for the onset of the propeller mode. This suggests that there could be a subset of fast-spinning WDs in more normal CV systems that are in a state of quiescence preceding a magnetic propeller phase. An example may be V1460~Her, for which the WD appears to have recently spun up, but does not yet possess a significant magnetic field. The lack of a significant field in V1460~Her could result from either its WD not reaching the onset of crystallization when the spin-up occurred, or that its field has not yet diffused to the surface.

\subsection{Differences Between AE~Aqr and J0240}

AE Aqr is more luminous than J0240 in the FUV. Outside of flares, the low-state continuum of AE~Aqr is consistently about 3.5 times more luminous than J0240. The increase in flux between their low state and high state is about a factor of 1.5 for both systems. However, the minimum luminosity of AE~Aqr is higher than that of J0240, so the factor of 1.5 increase in continuum flares results in significantly greater peak energy output than flaring in J0240. From our statistical analysis of ground-based transient surveys, flares in AE~Aqr are more luminous than in J0240 at optical wavelengths. Flares in AE~Aqr peak at more than 2.8 times the intrinsic brightness of J0240. 

The FUV continuum flux during flares is highly correlated with the emission line strength in AE~Aqr. For J0240, the emission line strengths vary significantly for the same continuum flux and may have an orbital dependence. This is particularly clear in the first half of orbit~2 ($0.15\leq\phi\leq0.22$) where the emission lines are weaker than over the eclipse. Optical depth in the lines may play an important role in determining total line emission as J0240 is a high inclination system with an outflow. This is made clear by the broad P-Cygni profile in \ion{Si}{4} near orbital phase 0.5.

Outside of eclipse, the difference in the minimum FUV continuum luminosity between AE Aqr and J0240 is likely explained by their WD properties including size, average temperature, and surface temperature distribution.

\section{Summary and Conclusions}

The FUV spectrum of J0240 shows the resonance lines characteristic of most CVs. While the \ion{N}{5}/\ion{C}{4} ratio in J0240 is anomalous relative to typical magnetic CVs, it is not as drastic as the inversion of the ratio in AE~Aqr, which is nearly devoid of \ion{C}{4} emission. The secondary in AE Aqr could have been stripped to more carbon-poor layers than the secondary of J0240. This may be due to a difference in the mass ratios between the two systems at the onset of mass transfer.

The presence of a high-velocity P-cygni-like absorption feature in the \ion{Si}{4} resonance line confirms the presence of a fast outflow. The orbital phase of the outflow is consistent with the propeller model \citep{Wynn1997}. At optical wavelengths, the outflow is detected as a narrow H$\alpha$ absorption \citep{Garnavich2021} feature, while in the FUV the feature is broad. The broad \ion{Si}{4} absorption is likely formed close to the launching of the propeller. 

The eclipse of the WD is observed in the UV during $HST$ orbit 1, and no flares are observed during the eclipse. This is consistent with optical observations \citep{Littlefield2020} and suggests that the flaring region is close to the WD. Emission lines continue to be detected during the eclipse, demonstrating that the lines are coming from a more extended location than the continuum flares.

The UV spin signal of the WD is detected and has a measured period of 24.939~$\pm$~0.006~s, which is consistent with the optical spin period found in \citet{Pelisoli2022}. The spin signal is not detected within the emission lines or during the eclipse. Outside of the eclipse, the significance of the detection of the spin signal varies with time. This could be due to the optical depth of the outflow, which may obscure the surface of the WD. Since the $HST$ observations only cover one binary orbit, it is unclear if the modulation of the spin amplitude is correlated with orbital phase, or varies randomly.

We constrain the mean temperature of the WD to be no more than about 11,000 to 15,000~K for a WD between 0.7 and 1.3~M$_{\odot}$. These temperature limits are generally consistent with those required for a WD to begin crystallization at its core. Cooler and older WDs are commonly observed to be magnetic \citep{Bagnulo2022}, and crystallization may be a prerequisite for the development of their magnetism, particularly in CVs \citep{Schreiber2021}.

J0240 and AE~Aqr show stark differences in their continuum and emission luminosities. While both systems flare, J0240 experiences less luminous flares than AE~Aqr, both in the UV continuum and emission lines. This difference is also seen in the optical continuum. If the two systems have similar mass WDs, then the difference in the low-state continuum luminosities in J0240 and AE Aqr is likely due a cooler average surface temperature on the J0240 WD.

The reason flares from J0240 are significantly weaker than from AE~Aqr in both the FUV and optical is not apparent. The WD in J0240 spins 30\%\ faster than its counterpart in AE~Aqr, but differences in magnetic field strength or mass transfer characteristics may be the cause of the difference in flare luminosities. 

There are now two CV propellers. They show significant differences in their FUV and optical properties. We interpret J0240 and AE~Aqr as possible members of a short-lived stage that precedes their evolution into a more standard IP. Although difficult, we expect that long-term monitoring of the spin signal in J0240 will eventually detect a spin-down similar to what is seen in AE~Aqr. This opens up a broad spectrum of opportunities to further develop theoretical frameworks that connect binary evolution with magnetohydrodynamics.

\begin{acknowledgments}

We thank J. Halpern and J. Thorstensen. This research was partially supported by STScI Program GO-16881 through NASA contract NAS5-26555.

This work is partly based on observations obtained at the MDM Observatory, operated by Dartmouth College, Columbia University, Ohio State University, Ohio University, and the University of Michigan.

This work has made use of data from the European Space Agency (ESA) mission
{\it Gaia} (\url{https://www.cosmos.esa.int/gaia}), processed by the {\it Gaia}
Data Processing and Analysis Consortium (DPAC,
\url{https://www.cosmos.esa.int/web/gaia/dpac/consortium}). Funding for the DPAC
has been provided by national institutions, in particular the institutions
participating in the {\it Gaia} Multilateral Agreement.

\end{acknowledgments}

\vspace{5mm}
\facilities{ $HST$ (COS), McGraw-Hill, ASAS-SN, ATLAS}

\software{astropy \citep{Astropy}}

\end{document}